\documentclass[prl,twocolumn,superscriptaddress,showpacs]{revtex4}
\usepackage{slashbox}
\usepackage{amsmath,graphicx,amsfonts,bm,amssymb}
\usepackage{times}
\begin{document}

\title{Quantum computation in a decoherence-free subspace with superconducting devices}

\author{Z.-Y. Xue\footnote{email: xuezhengyuan@yahoo.com.cn}}

\affiliation{Department of Physics and Center of Theoretical and
Computational Physics, The University of Hong Kong, Pokfulam Road,
Hong Kong, China}

\author{S.L. Zhu}

\affiliation{Department of Physics and Center of Theoretical and
Computational Physics, The University of Hong Kong, Pokfulam Road,
Hong Kong, China}

\affiliation{Laboratory of Quantum Information Technology, ICMP and
SPTE, South China Normal University, Guangzhou 510006, China}
\author{Z.D. Wang}
%\email{zwang@hkucc.hku.hk}

\affiliation{Department of Physics and Center of Theoretical and
Computational Physics, The University of Hong Kong, Pokfulam Road,
Hong Kong, China}

\date{\today}

\begin{abstract}
We propose a scheme to implement quantum computation in
decoherence-free subspace  with superconducting devices inside a
cavity by unconventional geometric manipulation. Universal
single-qubit gates in encoded qubit can be achieved with cavity
assisted interaction. A measurement-based two-qubit Controlled-Not
gate is produced with parity measurements assisted by an auxiliary
superconducting device and followed by prescribed single-qubit
gates. The measurement of currents on two parallel devices can
realize a projective measurement, which is equivalent to the parity
measurement on the involved devices.
\end{abstract}



\maketitle

%\section{Introduction}

Physical implementation of quantum computers relies on coherent and
accurate evolution to achieve quantum logical gates. Recently,
superconducting devices have attracted significant interest for the
hardware implementation of quantum computer because of their
potential scalability \cite{Makhlin}. In addition, the cavity
assisted interaction has been experimentally illustrated to have
several practical advantages \cite{yale}. But, decoherence and
systematic errors always occur in real quantum systems and therefore
stand in the way of physical implementation. Decoherence may quickly
destroy the information  stored in a quantum system. Indeed, it is
technically difficult for a single qubit survives for long on its
own. But by teaming up, a group of qubits can work together, forming
decoherence-free subspace (DFS) \cite{dfs}, to eliminate the
influence of their environment, and thus keeping their integrity.
For superconducting devices,  the short dephasing time  poses one of
main challenges in coherent controls, and thus it is significant to
figure out methods of improvement. To manipulate the quantum state,
one will also inevitably encounter systematic errors. Fortunately,
geometric manipulation of quantum information could result in
quantum gates that are robust against stochastic control errors
\cite{zhuerror}. Combination of the resilience of the DFS approach
against the environment-induced decoherence and the operational
robustness of geometric manipulation was also proposed with trapped
ions \cite{gd1,cen} and by engineering the environment \cite{gd2}.

In this paper, we work out a feasible scheme to implement quantum
computation based on DFS encoding with an extended unconventional
geometric scenario \cite{cen,Leibfried,zhuunconvetional,Zheng}. We
illustrate our idea by incorporating the superconducting devices
inside a cavity. Universal single-qubit gates in an encoded qubit
\cite{note} can be achieved with the help of cavity assisted
interaction. In particular, the realization of superconducting
parity measurements on two devices, together with single-device
measurements and single-qubit gates, is able to generate a two-qubit
Controlled-Not (CNOT) gate \cite{loss}. In this sense, this scheme
is the measurement-based quantum computation. The easy combination
of individual addressing and selective interaction with the
many-device setup proposed in the system presents a distinct merit
for physical implementation.

%\section{Implementation of the model}

%\subsection{The superconducting device}

\begin{figure}[bp]
\centering
\includegraphics[width=4.5cm]{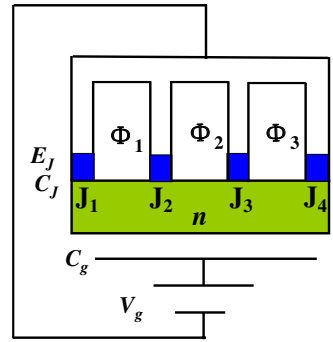}
\caption{Schematic illustration of the superconducting device as the
effective spin. Device made of two SQUIDs with a common
superconducting charge box. This more flexible design will introduce
more control variables of the effective spin.} \label{qubit}
\end{figure}

A  device for engineering the wanted interaction is shown in Fig.
\ref{qubit}. It consists of two superconducting quantum interference
devices (SQUIDs) with a common superconducting charge box that  has
$n$ excess Cooper-pair charges. Each SQUID is formed by two small
identical Josephson junctions (JJs) with the capacitance $C_{J}$ and
Josephson coupling energy $E_J$, pierced by an external magnetic
flux $\Phi_k$. A control gate voltage $V_g$ is connected to the
system via a gate capacitor $C_g$. $J_l$ with $l\in\{1, 2, 3, 4\}$
denotes the $l$th JJ. The gauge-invariant phase difference
$\varphi_l$ of $J_l$ is determined from the flux quantization for
the three independent loops, i.e.,
$\varphi_k-\varphi_{k+1}=2\pi\Phi_k/\phi_0\equiv2\phi_k$ with
$k\in\{1, 2, 3\}$ and $\phi_0=h/2e$ being the  flux quantum. Since
we here focus on the charge regime, a convenient basis we choose is
formed by the charge states, parameterized by the number of Cooper
pairs $n$ on the box with its conjugate
$\varphi=\sum_{l}\varphi_{l}/4$. At temperatures much lower than the
charging energy and restricting the gate charge to the range of
$\bar{n}\in [0,1]$, only a pair of adjacent charge states
$\{|0\rangle,|1\rangle\}$ on the island are relevant. Setting
$\phi_1=\phi_3=0$, the device Hamiltonian reduces to \cite{Makhlin}
\begin{equation}
\label{h1r} H_s=-E_{ce}\sigma_z-E_{\Phi}\sigma_x,
\end{equation}
where $E_{\Phi}=2E_J \cos\phi_2$ and $E_{ce} =2E_c(1-2\bar{n})$
with $E_{c}=e^2/2(C_g+4C_J)$ the charging energy and $\bar{n}=C_g
V_g/2e$  the induced charge controlled by the gate voltage $V_g$.

%\subsection{The switch method}

To produce the wanted interaction among devices, they are placed in
a cavity, being parallel to the plane perpendicular to the magnetic
component of the cavity mode, so that the cavity mode contributes an
additional component to the total magnetic flux as
$\varphi_{t}-\varphi_{t+1}=2\phi_t+g_t(a+a^{\dagger})\equiv2\tilde{\phi}_t$,
with $t\in\{1, 2, 3\}$ and $a$ ($a^\dagger$) as the creation
(annihilation) operator for the cavity mode. Devices are also placed
at the antinodes of the cavity mode and the size of the device is
negligible in comparison with the cavity mode wave length, so that
the device-cavity interaction constants $g_t$ of different devices
can be treated as the same one. For simplicity, we consider only the
single-mode standing wave cavity scenario, then the Hamiltonian
(\ref{h1r}) for a superconducting device in a cavity becomes
\begin{eqnarray}\label{hc1}
H_c&=&-E_{ce}\sigma_z-E_J\sum_{l=1}^4\cos\varphi_l\notag\\
&=&-E_{ce}\sigma_z-2E_J\bigg[\cos\tilde{\phi}_1\cos\left(\frac{\varphi_1+\varphi_2}{2}\right)\notag\\
&&+\cos\tilde{\phi}_3\cos\left(\frac{\varphi_3+\varphi_4}{2}\right)\bigg]\notag\\
&=&-E_{ce}\sigma_z-E_J\bigg\{\left(\cos\tilde{\phi}_1+\cos\tilde{\phi}_3\right)\notag\\
&&\qquad\times\bigg[\cos\left(\frac{\varphi_1+\varphi_2}{2}\right)+
\cos\left(\frac{\varphi_3+\varphi_4}{2}\right)\bigg]\notag\\
&&+\left(\cos\tilde{\phi}_1-\cos\tilde{\phi}_3\right)\notag\\
&&\qquad\times \bigg[\cos\left(\frac{\varphi_1+\varphi_2}{2}\right)-
\cos\left(\frac{\varphi_3+\varphi_4}{2}\right)\bigg]\bigg\}\notag\\
&=&-E_{ce}\sigma_z-2E_J\bigg[\left(\cos\tilde{\phi}_1+\cos\tilde{\phi}_3\right)
\cos\varphi\cos\theta\notag\\
&&+\left(\cos\tilde{\phi}_1-\cos\tilde{\phi}_3\right)
\sin\varphi\sin\theta\bigg],
\end{eqnarray}
where $\theta=(\varphi_1+\varphi_2-\varphi_3-\varphi_4)/4
=(\phi_1+2\phi_2+\phi_3)/2+(g_1+2g_2+g_3)(a+a^{\dagger})/4$ with
$\phi_1$ and $\phi_3$ being dc magnetic fluxes. Defining
$g=(g_1+2g_2+g_3)/4$ and set $\phi_1=\phi_3=0$, then
$\theta=\phi_2+g(a+a^{\dagger})$. Up to the first order of $g$,
i.e., in Lamb-Dicke limit, Hamiltonian (\ref{hc1}) becomes
\begin{eqnarray}\label{hc2}
H_c&=&-E_{ce}\sigma_z-4E_J\cos\varphi\cos\theta \notag\\
&\simeq& H_s+2gE_J\sin\phi_2(a+a^{\dagger})\sigma_x.
\end{eqnarray}
We can see that the interaction can be switched off by modulating
the external magnetic field as $\Phi_2=k\phi_0$ with $k$ an integer.
In other words, the qubit and the cavity evolve independently in
this case. The external flux is merely used to separately address
the qubit rotations, while the evolution of the qubit is governed by
Hamiltonian (\ref{h1r}) with the coefficient $E_{\Phi}$ being
replaced by $2E_J$.

%\subsection{The implementation}

In Ref. \cite{xuewangzhu}, it was assumed that the inter-SQUID
loop (enclosed by the flux $\Phi_2$ in Fig. \ref{qubit}) is much
larger than other two SQUID loops (enclosed by the fluxes $\Phi_1$
or $\Phi_3$), and thus neglected the cavity mediated interaction
in those loops. This would require a larger device size, and may
make it more sensitive to noises. Here, we briefly elaborate that
the wanted interactions among selected devices may also be induced
without the loop size restriction imposed in Ref.
\cite{xuewangzhu}. If $N$ devices are located within a single-mode
cavity, to a good approximation, the whole system may be
considered as $N$ two-level systems coupled to a quantum harmonic
oscillator \cite{zhu}. Assuming the devices to work in their
degeneracy points, the cavity-device interaction is given by
\begin{eqnarray}\label{hint}
H_{int}=-2E_J\sum_{j=1}^N\bigg[\left(\cos\tilde{\phi}_1^j+\cos\tilde{\phi}_3^j\right)
\cos\varphi_j\cos\theta_j\notag\\
+\left(\cos\tilde{\phi}_1^j-\cos\tilde{\phi}_3^j\right)
\sin\varphi_j\sin\theta_j\bigg],
\end{eqnarray}
where we have assumed $E_J^j=E_J$ for simplicity. Assuming
$g^j_t=g$, up to the first order of $g$, Hamiltonian (\ref{hint})
becomes
\begin{eqnarray}\label{hint2}
H_{int}&\simeq&-2E_J\sum_{j=1}^N\bigg\{\cos\varphi_j\bigg\{\left(\cos\phi_1^j+\cos\phi_3^j\right)\notag\\
&&\qquad\times\left[\cos\phi_2^j+g\sin\phi_2^j(a+a^{\dagger})\right]\notag\\
&&+g\left(\sin\phi_1^j+\sin\phi_3^j\right)(a+a^{\dagger})\cos\phi_2^j\bigg\}\notag\\
&&+\sin\varphi_j\bigg\{\left(\cos\phi_1^j-\cos\phi_3^j\right)\notag\\
&&\qquad\times\left[\sin\phi_2^j+g\cos\phi_2^j(a+a^{\dagger})\right]\notag\\
&&+g\left(\sin\phi_1^j-\sin\phi_3^j\right)(a+a^{\dagger})\sin\phi_2^j\bigg\}\bigg\}.
\end{eqnarray}
Setting $\phi_2=\omega t$ for all the selected devices and in the
interaction picture with respect to
\begin{eqnarray}\label{h0}
H_0=\hbar\omega_c(a^\dagger a+\frac{1}{2}),
\end{eqnarray}
Hamiltonian (\ref{hint2}) becomes
\begin{eqnarray}\label{hint3}
H_{int}&\simeq&\frac{gE_J}{2}\sum_{j=1}^N\bigg\{\sigma^x_j\bigg[i\left(\cos\phi_1^j+\cos\phi_3^j\right)
(ae^{i\delta t}-a^{\dagger}e^{-i\delta t})\notag\\
&&\qquad-\left(\sin\phi_1^j+\sin\phi_3^j\right)(ae^{i\delta t}+a^{\dagger}e^{-i\delta t})\bigg]\notag\\
&&-\sigma^y_j\bigg[\left(\cos\phi_1^j-\cos\phi_3^j\right)
(ae^{i\delta
t}+a^{\dagger}e^{-i\delta t})\notag\\
&&\qquad-i\left(\sin\phi_1^j-\sin\phi_3^j\right)(ae^{i\delta
t}-a^{\dagger}e^{-i\delta t})\bigg]\bigg\}
\end{eqnarray}
under the rotating-wave approximation, i.e.,
$0<\delta=\omega-\omega_c\ll\omega_c$. If $\phi_1=\phi_3=k\pi$, the
cavity mediated interaction of Eq. (\ref{hint3}) reduce to
\begin{eqnarray}
\label{hintxx}
H_{int}^{x}=i\hbar\beta\left(a^{\dagger}e^{-\text{i}\delta
t}-ae^{\text{i}\delta t}\right)J_x,
\end{eqnarray}
where $\beta=gE_J/\hbar$ and $J_{x, y, z}=\sum_{j=1}^N\sigma_j^{x,
y, z}$. In the case of large detuning ($\delta\gg\beta$) or
periodical evolution ($\delta t=2k\pi$), the corresponding effective
Hamiltonian is given by \cite{xuewangzhu,zhu,zheng2,Sorensen}
\begin{eqnarray}
\label{hx} H_x=\hbar\chi J_x^2,
\end{eqnarray}
where $\chi=\beta^2/\delta$. If $\phi_3=\phi_1-\pi=k\pi$, then the
reduced  effective Hamiltonian is
\begin{eqnarray}
\label{hy} H_{y}=\hbar\chi J_y^2.
\end{eqnarray}
Note that the Hamiltonian  (\ref{hx}) and (\ref{hy}) are independent
on the number of devices, and can also be obtained by periodical
dynamic evolution \cite{zhu}. This cavity assisted collision type of
Hamiltonian was first proposed for two atoms in cavity QED
\cite{zheng2} with experimental verification in \cite{zheng2exp}.

We now elucidate how to achieve universal single-qubit rotation
\cite{cen}. We employ the pair-bit code by which the logical qubit
is encoded in a subspace $\{|\textbf{0}\rangle,
|\textbf{1}\rangle\}$ as
\begin{equation}
|\textbf{0}\rangle _i=|0\rangle _{i_1}\otimes |1\rangle _{i_2},
\quad |\textbf{1}\rangle _i=|1\rangle _{i_1}\otimes |0\rangle
_{i_2}, \label{enqubit}
\end{equation}
where $i=1,\cdots ,N/2$ indexes qubits of an array of $N$ devices.
Such an encoding is the well-known DFS \cite{dfs} against the
collective dephasing of the system-bath interaction. Let us denote
$X$, $Y$, and $Z$  as the three Pauli matrices of the encoded qubit
subspace. The evolution operator for two selected devices interact
with Hamiltonian in Eq. (\ref{hx}) is
\begin{eqnarray} \label{ux}
U_x(\gamma)&=&\exp\left[-2\text{i}\gamma
\left(1+\sigma^x_{i_1}\sigma^x_{i_2}\right)\right]\nonumber\\
&\sim&\exp\left(-2\text{i}\gamma\sigma^x_{i_1}\sigma^x_{i_2}\right)=\exp\left(-2\text{i}\gamma
X\right),
\end{eqnarray}
where $\gamma=\chi t$. If we set $\phi_1=\phi_3=k\pi$ in device
$i_1$ and $\phi_3=\phi_1-\pi=k\pi$ in device $i_2$, then the
reduced evolution operator for the two selected devices is
\begin{eqnarray} \label{uy}
U_y(\gamma)\sim\exp\left(-2\text{i}\gamma\sigma^x_{i_1}\sigma^y_{i_2}\right)=\exp\left(-2\text{i}\gamma
Y\right).
\end{eqnarray}
Certainly, (\ref{ux}) and (\ref{uy}) are non-commutable,
constructing the well-known universal single-qubit rotations.

%\section{CNOT gate}

\begin{figure}[tbp]
\includegraphics[width=8.5cm]{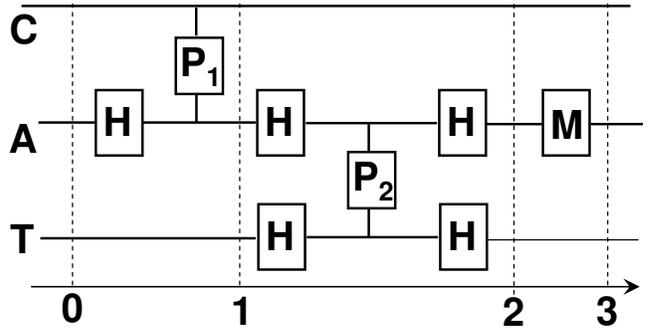}
\caption{Measurement-based CNOT gate for two encoded qubits. Capital
letters "C" and "T" represent the control and target qubit,
respectively. "A" represents an auxiliary device, it can witness the
qubit state via parity measurements "P", which operate on two
devices, one from "A" and the other from "C" or "T". "H" is the
Hadamard gate. The measurement "M" results of "A" in the
$\{|0\rangle, |1\rangle\}$ basis together with the outcomes of the
two parity measurements "P" determine which operation one has to
apply on the "C" and "T" qubit in order to complete the CNOT gate.
The arrowed line in the bottom represents the sequence of the
process. The point "0", "1", "2", and "3" stand for the initial
system state, the system states after measurements "$\text{P}_1$",
before and after "M", respectively. \label{cnot}}
\end{figure}

We next proceed to implement a CNOT  gate between two encoded qubits
with the help of an auxiliary device. Here we propose a
measurement-based CNOT gate operation \cite{loss}. The relevant
operations are single-qubit rotations, single-device
rotations/measurements, and effective parity measurements for two
devices. The circuit for the CNOT gate is depicted in Fig.
\ref{cnot}. The auxiliary device is initially prepared  in its
ground state $|0\rangle_A$.  The parity measurement is operated in
$\{|0\rangle, |1\rangle \}$ basis. The devices can be treated as
effective spin 1/2 systems, and the parity here represents for the
total spin for the two involved devices, which can be used to
witness the states of the involved spins \cite{loss}. After a
Hadamard gate on the auxiliary device, the first parity measurement
$\text{P}_1$ in Fig. \ref{cnot} is implemented on the auxiliary
device and the first device from "C" qubit.  After Hadamard rotation
of the auxiliary devices and the target qubit, the second parity
measurement $\text{P}_2$ is implemented on the auxiliary device and
the first device in the "T" qubit. Then we rotate back the auxiliary
device and the target qubit state by Hadamard gate. The last step is
the measurement of the auxiliary device in the $\{|0\rangle,
|1\rangle\}$ basis. The two parity measurement results, together
with the measurement result of the auxiliary device determine which
single-qubit gates to be operated on the control and target qubits
to generate a CNOT gate. The relationship between the measurement
results and the gates to be operated is summarized in the table
\ref{gate}. After completing the required gates on the corresponding
qubits, it is straightforward to check that the process is  a CNOT
gate operation between the two qubits.

\begin{table}
\centering \caption{Table of the correspondence between the
measurement results and the gates operated on the control and target
qubits. "0" and "1" represent odd and even parity, respectively.}
\label{gate}
\begin{tabular}{  c  c  c  c  c } \hline \hline
"P$_1$" & "P$_2$"  & result of "M" & gate on "C"  & gate on "T"\\
\hline
  1  &  1  &  $|0\rangle_A$  &  I  &  I \\ \hline
  1  & 1  &  $|1\rangle_A$  &  I  & X \\ \hline
 1  &  0  & $|0\rangle_A$  &  Z  &  I \\ \hline
  1  &  0 & $|1\rangle_A$  &  Z  &  X \\ \hline
  0  &  1  &  $|0\rangle_A$ &  I  &  X \\ \hline
  0  &  1  &  $|1\rangle_A$  &  I  & I \\ \hline
  0  &  0  &  $|0\rangle_A$ &  Z &  X \\ \hline
  0  &  0  &  $|1\rangle_A$  &  Z  &  I \\
  \hline \hline
\end{tabular}
\end{table}

%\section{An example}

To verify that a CNOT gate is implemented after the circuit
plotted in Fig. \ref{cnot}, we consider that the two qubits are
initially in the states
\begin{subequations}\label{ct}
\begin{equation}
|\psi\rangle_C=\left(\alpha|\bf{0}\rangle+\zeta|\bf{1}\rangle\right)_C,
\end{equation}
\begin{equation}
|\psi\rangle_T=\left(\xi|\bf{0}\rangle+\tau|\bf{1}\rangle\right)_T,
\end{equation}
\end{subequations}
where $|\alpha|^2+|\zeta|^2=1$ and $|\xi|^2+|\tau|^2=1$. The initial
state of the system at point 0 in Fig. \ref{cnot} is given by
\begin{equation}\label{1}
|\psi\rangle_C\otimes|0\rangle_A\otimes|\psi\rangle_T.
\end{equation}
The circuit in Fig. \ref{cnot}, together with prescribed
single-qubit gates, is to ensure the final state to be
\begin{eqnarray}\label{5}
\alpha|\bf{0}\rangle_C\left(\xi|\bf{0}\rangle
+\tau|\bf{1}\rangle\right)_T
+\zeta|\bf{1}\rangle_C\left(\xi|\bf{1}\rangle
+\tau|\bf{0}\rangle\right)_T,
\end{eqnarray}
up to a global phase. For the sake of definitiveness, let us single
out one of the possibilities as an example. If $\text{P}_1=0$,  the
system state at point 1 reduces to
\begin{eqnarray}\label{2}
\left(\alpha|\textbf{0}\rangle_C|1\rangle_A
+\zeta|\textbf{1}\rangle_C|0\rangle_A\right)\otimes|\psi\rangle_T.
\end{eqnarray}
If $\text{P}_2=1$, the system state at point 2 is
\begin{eqnarray}\label{3}
&&\frac{1}{2}\{\alpha|\bf{0}\rangle_C[(\tau+\xi)
(|\bf{0}\rangle+|\bf{1}\rangle)_T\otimes|\psi\rangle_A\nonumber\\
&&\qquad+(\tau-\xi) (|\bf{0}\rangle-|\bf{1}\rangle)_T\otimes\bar{|\psi\rangle}_A]\nonumber\\
&&+\zeta|\bf{1}\rangle_C[(\xi+\tau)(|\bf{0}\rangle+|\bf{1}\rangle)_T
\otimes|\psi\rangle_A\nonumber\\
&&\qquad+(\xi-\tau)(|\bf{0}\rangle-|\bf{1}\rangle)_T\otimes\bar{|\psi\rangle}_A]\}.
\end{eqnarray}
where $|\psi\rangle_A=(|0\rangle+|1\rangle)_A/\sqrt{2}$ and
$\bar{|\psi\rangle}_A=(|0\rangle-|1\rangle)_A/\sqrt{2}$. If the
measurement result of the auxiliary devices is $|0\rangle_A$, the
system state at point 3 is
\begin{eqnarray}\label{4}
\alpha|\bf{0}\rangle_C(\tau|\bf{0}\rangle
+\xi|\bf{1})_T+\zeta|\bf{1}\rangle_C(\xi|\bf{0}\rangle
+\tau|\bf{1}\rangle)_T,
\end{eqnarray}
which relates to the targeted final state (\ref{5}) up to a X-gate
on the target qubit (c.f. the table). Thus a nontrivial two-qubit
CNOT gate is achieved.

%\section{The superconducting parity meter}

\begin{figure}[tbp]
\includegraphics[width=8.5cm]{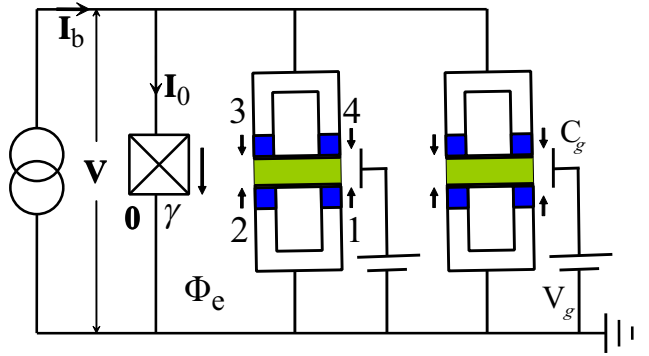}
\caption{A Josephson-Junction circuit with one large junction "0"
and two parallel charge devices. One of the devices is from the
encoded qubit and the other is its auxiliary device. Each device
consists of two SQUID loops. The small arrow near each JJ denotes
the direction of its phase drop. $\Phi_e$ is the dc external
magnetic flux of the loop consists of junction "0" and the first
device, which are related to the inter-SQUID magnetic flux of the
devices, and the cavity mediated interaction can be neglected in
this situation. The external magnetic flux of the SQUID loops in
both devices are set to be zero during the parity measurement.
\label{parity}}
\end{figure}

At this stage, we elaborate  how to implement a parity meter for
superconducting devices \cite{vion,you,wangxb}. Let us consider a
circuit with one large junction denoted by "0" and two parallel
devices ($c$ and $t$) made up of smaller JJs, as shown in Fig.
\ref{parity} \cite{you}. Under an external bias current $I_b$, the
current flowing through the large junction may be written as
\begin{eqnarray}
I_0=|I_b+I_d|=\left|I_b+\langle \psi_{1, 2}|\hat I |\psi_{1,
2}\rangle\right|
\end{eqnarray}
where $\hat I$ is the current operator for the two parallel devices
and $I_d$ is the sum of their expectation values. If $I_0>I_{c}$
with $I_c$ as the critical current of the large junction, the large
junction is switched from the superconducting state (with zero
voltage across the junction) to the normal state (with a nonzero
voltage $V$). As $\hat I$ is related to the device's state, by
monitoring the voltage across the junction one can determine which
type of state those JJ devices have been projected to \cite{vion},
and thus realize a quantum-state selector \cite{you,wangxb} (see
below for details). If $I_b$ is set to be significantly smaller than
$I_c$ and given the fact that $I_d\ll I_b$, then $I_0$ will always
be less than $I_c$, i.e., no measurement is in effect. Therefore, by
a proper choice of the bias current $I_b$, we are able to realize
effectively switching on/off of the process.

It is notable that the device in Fig. \ref{parity} is the same as
that of in Fig. \ref{qubit}. In Fig. \ref{parity}, we have chosen
the magnetic flux of SQUID loops, $\Phi_1$ and $\Phi_3$ in Fig.
\ref{qubit}, to be zero in each device, which simplifies our
calculation \cite{you}. With such choice, the constrain of the
inter-SQUID loop for each device is
$\varphi_2-\varphi_3=\varphi_1-\varphi_4=2\pi\Phi_e/\phi_0-\gamma$,
i.e., $\Phi_2\equiv 2\pi\Phi_e/\phi_0-\gamma$ for both devices,
where $\gamma$ is the gauge phase drop of the large JJ. For the
two-device case, setting $\Phi_{e}=\phi_0/2$, the total current
operator of both parallel devices is given by \cite{you}
\begin{eqnarray}\label{I}
\hat I=I_{1}\sigma_x^1+I_{2}\sigma_x^2,
\end{eqnarray}
which is state-dependent with $I_{1(2)}$  being the critical current
of the SQUID in device $1(2)$. To implement the parity measurement,
we choose $I_b= I_{c}-(I_{1}+I_{2})/2$ \cite{wangxb}, i.e.,
\begin{eqnarray}
I_0= I_{c}-(I_{1}+I_{2})/2+\langle \psi_{1, 2}|\hat I |\psi_{1,
2}\rangle.
\end{eqnarray}
Denote states $|\pm\rangle$ as the eigenstates of $\sigma_x$ with
eigenvalues $\pm1$, i.e., $\sigma_x|\pm\rangle=\pm|\pm\rangle$. If
$\psi_{1, 2}=|+\rangle_1|+\rangle_2$, then
\begin{eqnarray}
I_0= I_{c}+{I_{1}+I_{2} \over 2}
>I_c,
\end{eqnarray}
therefore the large junction is switched from the superconducting
state to the normal state with a nonzero voltage $V_1$. For the
other three  cases $\psi_{1, 2}\in\{|+\rangle_1|-\rangle_2$,
$|-\rangle_1|+\rangle_2$, $|-\rangle_1|-\rangle_2\}$, it is direct
to check $I_0 < I_c$. In other words, if $V_1\neq0$, the projective
measurement
\begin{eqnarray}
P_1^\prime=|+\rangle_1|+\rangle_{2,2}\langle+|_1\langle+|
\end{eqnarray}
is implemented on the two involved devices. For $V_1=0$, we may
reverse both the external field $\Phi_e$ and bias current $I_b$ to
their opposite directions, and  monitor the voltage again. If
$V_2\neq0$, then
\begin{eqnarray}
P_2^\prime=|-\rangle_1|-\rangle_{2,2}\langle-|_1\langle-|
\end{eqnarray}
is implemented. If $V_2=0$ again, this corresponds to the
measurement
\begin{eqnarray}
P_3^\prime=|+\rangle_1|-\rangle_{2,2}\langle+|_1\langle-|+|-\rangle_1|+\rangle_{2,2}\langle-|_1\langle+|.
\end{eqnarray}
It is obvious that $P_1^\prime$ and $P_2^\prime$ are even parity,
while $P_3^\prime$ is odd parity. This constructs a superconducting
parity meter in the $\{|\pm\rangle\}$ basis. Rotation of the device
state before and after the measurement results in the parity meter
in the $\{|0\rangle, |1\rangle\}$ basis, which is adopted in our
implementation of the CNOT gate.  It is also needed to measure the
auxiliary devices in the present implementation of the CNOT gate,
which can also be achieved with a minor modification of the setup
\cite{you}.

We now briefly address the experimental feasibility of our scheme.
Individual addressability is normally a prerequisite in any quantum
manipulation. Here, the size of the device setup is macroscopic,
thus individual addressability is taken as granted. Meanwhile, local
controllability of single qubit is obtained by conventional methods
\cite{Makhlin}. The cavity-device coupling and decoupling can be
controlled by the external magnetic flux, which can be effectively
controlled. This also ensures the selective cavity-device
interaction. In addition, the implementation set the devices working
in their degeneracy points, where they possess long coherence time
and minimal charge noises. Typical gate operation time is $t\sim10$
ns \cite{zhu}, which is much shorter than both the lifetime of qubit
and cavity decay time (at least on the order of $\mu$s
\cite{Makhlin,zheng2}). Imperfect control of time results in the
fluctuation of periodical condition while cavity decay forbids the
cavity state back to the original point in phase space, this
contribute to the decoherence in current implementation. However,
detailed examinations \cite{zheng3,chen,song} show that  these will
only result a little bit infidelity of the gate operation.

%\section{Conclusion}

In summary, we have proposed a feasible scheme to implement quantum
computation in the DFS  with superconducting devices inside a
cavity. The wanted interaction between selective devices can be
implemented. Universal single-qubit gates can be achieved with
cavity assisted interaction. A measurement-based two-qubit CNOT gate
is produced with parity measurements assisted by an auxiliary device
and followed by prescribed single-qubit gates. The easy combination
of individual addressing and selective interaction with the
many-device setup proposed in the system presents a distinct merit
for our physical implementation.

\bigskip

This work was supported by the RGC of Hong Kong under Grants Nos.
HKU7045/05P and HKU7049/07P plus HKU7044/08P, the NSFC under Grants
No. 10429401 and No. 10674049, and the State Key Program for Basic
Research of China (No. 2006CB921800 and No. 2007CB925204).


\begin{thebibliography}{99}


\bibitem{Makhlin} J. Q. You and F. Nori, Phys.
Today  {\bf 58} (11) (2005) 42;  Y. Makhlin, G. Sch\"{o}n, and A.
Shnirman, Rev. Mod. Phys. {\bf 73} (2001)  357 .

\bibitem{yale} A. Wallraff, D. I. Schuster, A. Blais, L. Frunzio, R.-S. Huang, J.
Majer, S. Kumar, S. M. Girvin, and R. J. Schoelkopf, Nature (London)
\textbf{431} (2004)  162.

\bibitem{dfs} L.-M. Duan and G. C. Guo, Phys. Rev. Lett. \textbf{79} (1997)  1953; P.
Zanardi and M. Rasetti,  Phys. Rev. Lett. \textbf{79} (1997) 3306;
D. A. Lidar, I. L. Chuang, and K. B. Whaley,  Phys. Rev. Lett.
\textbf{81} (1998) 2594 .


\bibitem{zhuerror} S.-L. Zhu and P. Zanardi, Phys. Rev. A \textbf{72} (2005)  020301(R).

\bibitem{gd1} L.-A. Wu, P. Zanardi, and D. A. Lidar,
Phys. Rev. Lett. \textbf{95} (2005)  130501; X. D. Zhang, Q. H.
Zhang, and Z. D. Wang, Phys. Rev. A \textbf{74} (2006)  034302.


\bibitem{cen} L.-X. Cen, Z. D. Wang, and S. J. Wang, Phys. Rev. A \textbf{74}
(2006)  032321.

\bibitem{gd2} A. Carollo, M. Fran\c{c}a Santos, and V. Vedral, Phys. Rev. Lett.
\textbf{96} (2006)  020403; Z.-Q. Yin, F.-L. Li, and P. Peng, Phys.
Rev. A \textbf{76} (2007)  062311.

\bibitem{Leibfried} D. Leibfried, B. DeMarco, V. Meyer, D. Lucas, M. Barrett, J. Britton,
W.M. Itano, B. Jelenkovic, C. Langer, T. Rosenband, and
D.J.Wineland, Nature (London) {\bf 422} (2003)  412.


\bibitem{zhuunconvetional} S.-L. Zhu and Z.D. Wang, Phys. Rev. Lett. {\bf
91}  (2003) 187902; J. Du, P. Zou, and Z. D. Wang, Phys. Rev. A {\bf
74} (2006) 020302.


\bibitem{Zheng} S. B. Zheng, Phys. Rev. A {\bf 70} (2004) 052320;
 Phys. Rev. A  {\bf 74}  (2006) 032322.



\bibitem{note} In order to avoid confusion, we use "qubit" to denote the
encoded logical qubit, while "device" for the superconducting
charge qubit schemeticed in Fig. 1 through out this paper.



\bibitem{loss} O. Zilberberg, B. Braunecker, and D. Loss, Phys. Rev. A
\textbf{77} (2008) 012327.


\bibitem{xuewangzhu}  Z.-Y. Xue, Z. D. Wang, and S.-L. Zhu, Phys. Rev. A
\textbf{77} (2008) 024301.

\bibitem{zhu} S.-L. Zhu, Z. D. Wang, and P. Zanardi, Phys. Rev. Lett.
{\bf 94} (2005) 100502; Z.-Y. Xue and Z. D. Wang, Phys. Rev. A
\textbf{75} (2007) 064303.


\bibitem{Sorensen} A. S{\o}rensen and K. M{\o}lmer, Phys. Rev. A {\bf 62} (2000) 022311.

\bibitem{zheng2} S.-B. Zheng and G.-C. Guo, Phys. Rev. Lett. \textbf{85} (2000)
2392;  S.-B. Zheng, Phys. Rev. A \textbf{66}  (2002) 060303(R).


\bibitem{zheng2exp} S. Osnaghi, P. Bertet, A. Auffeves, P. Maioli, M. Brune,
J. M. Raimond, and S. Haroche, Phys. Rev. Lett. \textbf{87}, 037902
(2001).


\bibitem{vion} D. Vion,  A. Aassime, A. Cottet, P. Joyez, H. Pothier, C. Urbina,
D. Esteve, and M. H. Devoret, Science {\bf 296} (2002) 886.

\bibitem{you} J. Q. You, J. S. Tsai, and F. Nori, Phys. Rev. Lett. {\bf 89} (2002)
197902; Phys. Rev. B {\bf 68} (2003) 024510.

\bibitem{wangxb} X.-B. Wang, J. Q. You, and F. Nori,  Phys. Rev. A \textbf{77} (2008)
062339; quant-ph/0608205 (2006).

\bibitem{zheng3} S.-B. Zheng, Phys. Rev. Lett.  \textbf{95} (2005) 080502.

\bibitem{chen} C.-Y. Chen et al., Phys. Rev. A \textbf{73} (2006) 032344; ibid.
\textbf{74} (2006) 032328.

\bibitem{song} Z.-G. Shi, X.-W. Chen, and K.-H. Song, J. Phys. B: At.
Mol. Opt. Phys. \textbf{42} (2009) 035504.

\end{thebibliography}
\end{document}